\documentstyle[preprint,eqsecnum,aps]{revtex}

\begin{document}
\draft
\title{A model of the Universe including Dark Energy accounted for by both a Quintessence Field and a (negative) Cosmological Constant}
\date{\today}
\author{Rolando Cardenas\thanks{rcardenas@mfc.uclv.edu.cu}, Tame Gonzalez\thanks{tame@mfc.uclv.edu.cu}, Yoelsy Leiva\thanks{yoelsy@mfc.uclv.edu.cu}, Osmel Martin\thanks{osmel@mfc.uclv.edu.cu}and Israel Quiros\thanks{israel@mfc.uclv.edu.cu}}
\address{Departamento de Fisica. Universidad Central de Las Villas. Santa Clara. CP: 54830 Villa Clara. Cuba}
\maketitle

\begin{abstract}
In this work we present a model of the universe in which dark energy is modelled explicitely with both a dynamical quintessence field and a cosmological constant. Our results confirm the possibility of a future collapsing universe (for a given region of the parameter space), which is necessary for a consistent formulation of string theory and quantum field theory. We have also reproduced the measurements of modulus distance from supernovae with good accuracy.  
\end{abstract}

\section{Introduction}

From 1998 to date several important discoveries in the astrophysical
sciences have being made, which have given rise to the so called New Cosmology \cite{turner1,turner2}. Amongst its more important facts we may cite: the universe expands in an acelerated way \cite{riess,perlmutter}; the first Doppler peak in the cosmic
microwave background is strongly consistent with a flat universe whose
density is the critical one \cite{cmb}, while several independent
observations indicate that matter energy density is about one third of the aforementioned critical density \cite{Om,smoot}. The last two facts implied that some
unknown component of the Universe ''was missing'', it was called dark
energy, and represents near two thirds of the energy density of the
universe. The leading candidates to be identified with dark energy involve
fundamental physics and include a cosmological constant (vacuum energy), a
rolling scalar field (quintessence), and a network of light, frustrated
topological defects \cite{mst}.

On the other hand, an eternally accelerating universe seems to be at odds
with string theory, because of the impossibility of formulating the
S-matrix. In a de Sitter space the presence of an event horizon, signifying
causally disconnected regions of space, implies the absence of asymptotic
particle states which are needed to define transition amplitudes \cite
{banks,cline}. This objection against accelerated expansion also applies to quantum field theory (QFT)\cite{sasaki}.

Due to the above there is a renewed interest in exponential quintessence,
because in several scenarios exponential potentials can reproduce the
present acceleration and predict future deceleration, so again string theory
has well defined asymptotic states \cite{cline,kl}. Worthwhile to notice
that exponential quintessence had been so far overlooked on fine tuning
arguments, but several authors have recently pointed out that the degree of
fine tuning needed in these scenarios is no more than in others usually
accepted \cite{cline,kl,rubano}.

The cosmological constant can be incorporated into the quintessence
potential as a constant which shifts the potential value, especially, the
value of the minimum of the potential, where the quintessence field rolls
towards. Conversely, the height of the minimum of the potential can also be
regarded as a part of the cosmological constant. Usually, for separating
them, the possible nonzero height of the minimum of the potential is
incorporated into the cosmological constant and then set to be zero. The
cosmological constant can be provided by various kinds of matter, such as
the vacuum energy of quantum fields and the potential energy of classical
fields and may also be originated in the intrinsic geometry. So far there is
no sufficient reason to set the cosmological constant (or the height of the
minimum of the quintessence potential) to be zero \cite{hwang}. In
particular, some mechanisms to generate a negative cosmological constant
have been pointed out \cite{ss,gh}.

The goal of this paper is to present a model of the universe in which the dark energy component is accounted for by both a quintessence field and a negative cosmological constant. The quintessence field accounts for the present stage of accelerated expansion of the universe. Meanwhile, the inclusion of a negative cosmological constant warrants that the present stage of accelerated expansion will be, eventually, followed by a period of collapse into a final cosmological singularity (AdS universe). 

\section{The Model}

Our scenario is a generalization of that of Rubano and Scudellaro \cite
{rubano}. We consider a model consisting of a three-component cosmological
fluid: matter, scalar field (quintessence with an exponential potential) and
cosmological constant. ''Matter'' means barionic + cold dark matter, with no
pressure, and the scalar field is minimally coupled and noninteracting with
matter. This model cannot be used from the very beginning of the universe,
but only since decoupling of radiation and dust. Thus, we don't take into
account inflation, creation of matter, nucleosynthesis, etc. Also, we use
the experimental fact of a spatially flat universe \cite{bernardis}. We
apply the same technique of adimensional variables we used in \cite{cmq} to
determine the integration constants without additional assumptions.

The action of the model under consideration is given by

\begin{equation}
S=\int d^4 x\;\sqrt{-g}\{\frac{c^2}{16\pi G}(R-2\Lambda)+{\cal L}_\phi+{\cal %
L}_{m}\},
\end{equation}
where $\Lambda$ is the cosmological constant, ${\cal L}_{m}$ is the
Lagrangian for the matter degrees of freedom and the Lagrangian for the
quintessense field is given by

\begin{equation}
{\cal L}_{\phi}=-\frac{1}{2}\phi_{,n} \phi^{,n}-V(\phi).
\end{equation}

\bigskip We use the dimensionless time variable $\tau =H_{0}t$, where $t$ is
the cosmological time and $H_{0}$ is the present value of the Hubble
parameter. In this case $a(\tau )=\frac{a(t)}{a(0)}$ is the scale factor. Then we have that, at present $(\tau =0)$

\begin{eqnarray}
a(0)&=&1,  \nonumber \\
\dot{a}(0)&=&1,  \nonumber \\
H(0)&=&1.
\end{eqnarray}

\bigskip Considering a spatially flat, homogeneous and isotropic universe,
the field equations derivable from (2.1) are

\begin{equation}
(\frac{\dot{a}}{a})^2=\frac{2}{9}\sigma^2 \{ \frac{\bar D}{a^3}+\frac{1}{2} 
\dot{\phi}^2 + \bar W(\phi) \},
\end{equation}

\begin{equation}
2\frac{\ddot{a}}{a}+(\frac{\dot{a}}{a})^2=-\frac{2}{3}\sigma^2 \{ \frac{1}{2}%
\dot{\phi}^2-\bar W(\phi)\},
\end{equation}
and

\begin{equation}
\ddot{\phi}+3\frac{\dot{a}}{a}\dot{\phi}+\bar{W}^{\prime}(\phi)=0,
\end{equation}
where the dot means derivative in respet to $\tau$ and, $\bar{W}(\phi)=\bar{V}(\phi)+\frac{3}{2}\frac{\bar{\Lambda}}{\sigma^2}$, $\bar{V}(\phi)=\bar{B}^2 e^{-\sigma\phi}$. The dimensionless constants are: $\bar{B}^{2}=\frac{B^{2}}{H_{0}^{2}}$, $\bar{\Lambda}=\frac{\Lambda }{H_{0}^{2}}$, $\bar{D}=\frac{D}{a_{0}^{3}H_{0}^{2}}=\frac{\rho _{m_{0}}}{H_{0}^{2}}$, with $\rho _{m_{0}}$ - the present density of matter, $\sigma ^{2}=\frac{12\pi G
}{c^{2}}$ and $B^{2}$ - a generic constant.

Applying the Noether Symmetry Aproach \cite{rmrs,r,crrs,rr}, it can be shown that the new variables we should introduce to simplify the
field equations are the same used in \cite{rubano}: $a^3=u v$ and $\phi=-\frac{1}{\sigma}\ln(\frac{u}{v})$.

In these variables the field Eqs. (2.4-2.6) may be written as the following pair of equations

\begin{equation}
\frac{\ddot{u}}{u}+\frac{\ddot{v}}{v}=\bar{B}^2\sigma^2\frac{u}{v}-\sigma^2%
\bar{V}_{0},
\end{equation}
and

\begin{equation}
\frac{\ddot{u}}{u}-\frac{\ddot{v}}{v}=-\sigma ^{2}\bar{B}^{2}\frac{u}{v},
\end{equation}
respectively. In Eq (2.7) we have introduced explicitely a negative
cosmological constant: $\bar{\Lambda}=-\frac{2}{3}\sigma^2\bar{V}_{0}$. Combining of Eqs. (2.7) and (2.8)yields:

\begin{equation}
\ddot{u}=-\frac{\sigma^2\bar{V}_{0}}{2}u,
\end{equation}

and

\begin{equation}
\ddot{v}=-\frac{\sigma^2\bar{\vee}_{0}}{2}v+\sigma^2\bar{B}^2u.
\end{equation}

The solutions of the equations (2.9) and (2.10) are found to be

\begin{equation}
u(\tau)=u_{1}\sin(\sigma\sqrt{\frac{\bar{V}_0}{2}}\tau)+u_{2}\cos(\sigma%
\sqrt{\frac{\bar{V}_0}{2}}\tau),
\end{equation}
and

\begin{eqnarray}
v(\tau ) &=&\{v_{2}+\frac{\bar{B}^{2}}{2\bar{V}_{0}}u_{2}-\frac{\sigma \bar{B%
}^{2}}{\sqrt{2\bar{V}_{0}}}u_{1}\tau \}\cos (\sigma \sqrt{\frac{\bar{V}_{0}}{%
2}}\tau )+  \nonumber \\
&+&\{v_{1}+\frac{\bar{B}^{2}}{2\bar{V}_{0}}u_{1}+\frac{\sigma \bar{B}^{2}}{%
\sqrt{2\bar{V}_{0}}}u_{2}\tau \}\sin (\sigma \sqrt{\frac{\bar{V}_{0}}{2}}%
\tau ),
\end{eqnarray}
where $u_{1},u_{2},v_{1}$ and $v_{2}$ are the integration constants. These can be related with the initial conditions for $u$ and $v$ (and their $\tau$-derivatives) thoughout the following equations,

\begin{eqnarray}
u(0)&=&u_2\;\; ,  \nonumber \\
\dot u(0)&=&\sigma\sqrt{\frac{\bar V_0}{2}}u_1\;\;,
\end{eqnarray}
and

\begin{eqnarray}
v(0)&=&v_2+\frac{\bar B^2}{2\bar V_0}u_2,  \nonumber \\
\dot v(0)&=&\sigma\sqrt{\frac{\bar V_0}{2}}(v_1-\frac{\bar B^2}{2\bar V_0}%
u_1).
\end{eqnarray}

In finding the integration constants we first evaluate Eq. (2.9) for $\tau =0$ and then we consider the relationships (2.3). Another expression is provided by
the Hubble parameter: $H(\tau)=\frac{1}{3}\{\frac{\dot u(\tau)}{u(\tau)}+\frac{\dot v(\tau)}{v(\tau)}\}$, evaluated at $\tau=0$. Other two relationships needed for our purposses are given by the field equations (2.4) and (2.5). We introduced the deceleration parameter: $q(\tau)\}=-\{1+\frac{\dot H(\tau)}{H(\tau)^2}\}$. We find 4 sets of solutions, or, strictly speaking, 4 branches of a same solution.

\begin{equation}
u_2^{(\pm)}=\pm\sqrt{\frac{3(2-q_0)-\sigma^2\bar D+2\sigma^2\bar V_0}{%
2\sigma^2\bar B^2}},
\end{equation}

\begin{equation}
v_2^{(\pm)}=\frac{1-\frac{\bar B^2}{2\bar V_0}u_2^2}{u_2^{(\pm)}},
\end{equation}

\begin{equation}
u_{1\;[\pm]}^{(\pm)}=\frac{\{3-[\pm]\sqrt{3(1+q_0)-\sigma^2\bar D}\}}{\sqrt{%
2\sigma^2\bar V_0}}u_2^{(\pm)},
\end{equation}
and

\begin{equation}
v_{1\;[\pm]}^{(\pm)}=\frac{6-\sqrt{2\sigma^2\bar V_0}\;v_2^{(\pm)}u_{1\;[%
\pm]}^{(\pm)}}{\sqrt{2\sigma^2\bar V_0}\;u_2^{(\pm)}},
\end{equation}
where $q_{0}$ is the present value of the deceleration parameter. If we introduce new parameters; $\Omega_{m_0}=\frac{2}{9}\sigma^2\bar D$, $\Omega_{Q_0}=\frac{2}{9}\sigma^2\{\frac{1}{2}\dot\phi(0)^2+\bar V(\phi(0))\}$, and $\Omega_{\Lambda}=-\frac{2}{9}\sigma^2\bar V_0$, where, for flat FRW spacetimes, $\Omega_{m_0}+\Omega_{Q_0}+\Omega_{\Lambda}=1$, and we propose the following relationship between the
parameters $\bar B^2 $ and $\bar V_0$:

\begin{equation}
\bar B^2=n\;\bar V_0,
\end{equation}
where $n$ is a positive real number, then the above integration constants
can be written in the following way:

\begin{equation}
u_2^{(\pm)}=\pm\sqrt{\frac{2-q_0-1.5\Omega_{m_0}-3\Omega_\Lambda}{%
-3n\Omega_\Lambda}},
\end{equation}

\begin{equation}
v_2^{(\pm)}=\frac{1-\frac{n}{2}\;u_2^2}{u_2^{(\pm)}},
\end{equation}

\begin{equation}
u_{1\;[\pm]}^{(\pm)}=\frac{\{\sqrt{3}-[\pm]\sqrt{1+q_0-1.5\Omega{m_0}}\}}{%
\sqrt{-3\;\Omega_\Lambda}}u_2^{(\pm)},
\end{equation}
and

\begin{equation}
v_{1\;[\pm]}^{(\pm)}=\frac{2-\sqrt{-\Omega_\Lambda}\;v_2^{(\pm)}u_{1\;[%
\pm]}^{(\pm)}}{\sqrt{-\Omega_\Lambda}\;u_2^{(\pm)}},
\end{equation}
respectively.

Since $\sqrt{1+q_{0}-1.5\Omega _{m_{0}}}$ should be real then, the following constrain on the present value of the deceleration parameter follows: $q_0\geq -1+1.5\Omega_{m_0}$.

\bigskip It can be noticed that the constants (and, consequently, the
solutions) depend on 4 physical parameters: $\Omega _{m_{0}}$, $\Omega
_{\Lambda }$, $q_{0}$ and on the positive real number $n$. In general,
the experimental values for $\Omega _{m_{0}}$ and $q_{0}$ are
model-dependent, because this magnitudes are not directly measured (though
Turner and Riess have developed a model-independent test for past
deceleration \cite{tr}). 

Concerning the parameter $n,$ it should be noticed that most of the relevant
cosmological parameters (such as the scale factor and the deceleration parameter) are quite insensitive to it's value. However, the state parameter is dependent on it. Though we made calculations for several values of $\ \Omega _{\Lambda }$ in the range
-0.01 a -0.30, for simplicity we present results for -0.15, having in
mind that they change little for other values.

\section{Analysis of Results}

After making a detailed study, it was determined that the only relevant cosmological magnitude that has a sensible dependence on parameters $n$ is the state parameter $\omega$. We fixed $\Omega_{m_{0}}=0.3$, in accordance with experimental evidence.
 
Figure 1 shows the evolution of the scale factor for $\Omega _{\Lambda
}=-0.15$. It was shown both
algebraically and graphically that the evolution of the universe is independent of $n$, but not writen
for the sake of simplicity. We also saw that with the decrease (modular increase)
of $\ \Omega _{\Lambda }$, the time of collapse diminishes.

We got the dependence of the state parameter and of
the deceleration parameter with the redshift and them we selected the values of $n$ and $q_{0}$. For this purpose we considered the results
of Turner and Riess\cite{tr}.

Figure 2 shows the behaviour of the deceleration parameter as function of the
redshift z for the
same values of the parameters. This figure shows an early stage of deceleration and a current epoch of acceleration. A transition from an accelerated phase
to a decelerated one is seen approximately for z =0.5. We appreciate an increase of
the deceleration parameter upon increasing the value of z. This points at a past epoch in the evolution when gravity of the dark energy was attractive. As follows from figure 1, aceleration is not eternal: in the future $q>0$ again,
which gives rise to the collapse.

Figure 3 shows the evolution of the state parameter of the effective quintessence field $\omega
_{\phi }$. It's noticeable that the effective quintessence field has state
parameter $\omega_{\phi }$ near $-1$ today, which means that its behaviour is similar to the ''pure'' cosmological constant, as a vacuum fluid. 

If we are to explain the
very desirable for today's cosmology recent and future deceleration obtained
in our model, it's important to look at the dynamical quintessence field. We
see that in the recent past $\omega_{\phi }>0$, which implies that
quintessence field behaved (or simply was) like ordinary atractive matter,
giving rise to the logical deceleration. In the future this will happen
again ($\omega_{\phi }>0)$ , with the consequent deceleration.

The present values
of the physical parameters ($\Omega _{m_{0}}=0.3$, $\Omega _{\Lambda }=-0.15$,
$q_{0}=-0.44$) were chosen after a detailed analisys of the behaviour of these parameters shown in figs. 2 and 3.

Now we proceed to analyze how our solution reproduces experimental results.
With this purpose, in Fig. 4 we plot the distance modulus $\delta (z)$ vs
redshift z, calculated by us and the one obtained with the usual model with
a constant $\Lambda $ term. The relative deviation is of about 0.5$\%$.

\section{Conclusions}

In a recent paper \cite{hwang} it is pointed out that the ultimate fate of the evolution of our
universe is much more sensitive to the presence of the cosmological constant
than any other matter content. In particular, the universe with a negative
cosmological constant will always collapse eventually, even though the
cosmological constant may be nearly zero and undetectable at all at the
present time. Our results support the very general
assertions of \cite{hwang}, we have shown that for a determined region of the parameter space, the universe collapses. This also favours the consistent formulation of string theory and quantum field theory, as explained in the introduction. 
The experimental measurements of modulus distance from the supernovae are adequately reproduced within an accuracy of 0.5$\%$. 
So far, we have investigated one of the several possible branches of the solution, leaving for the future the investigation of the others. We have also reserved for future work the careful examination of this universe near its beginning (i.e., just after the decoupling of matter and radiation).

We acknowledge Claudio Rubano, Mauro Sereno and Paolo Scudellaro, from Universita di Napoli "Federico II" , Italy, for useful comments and discussions and Andro Gonzales for help in the computations.

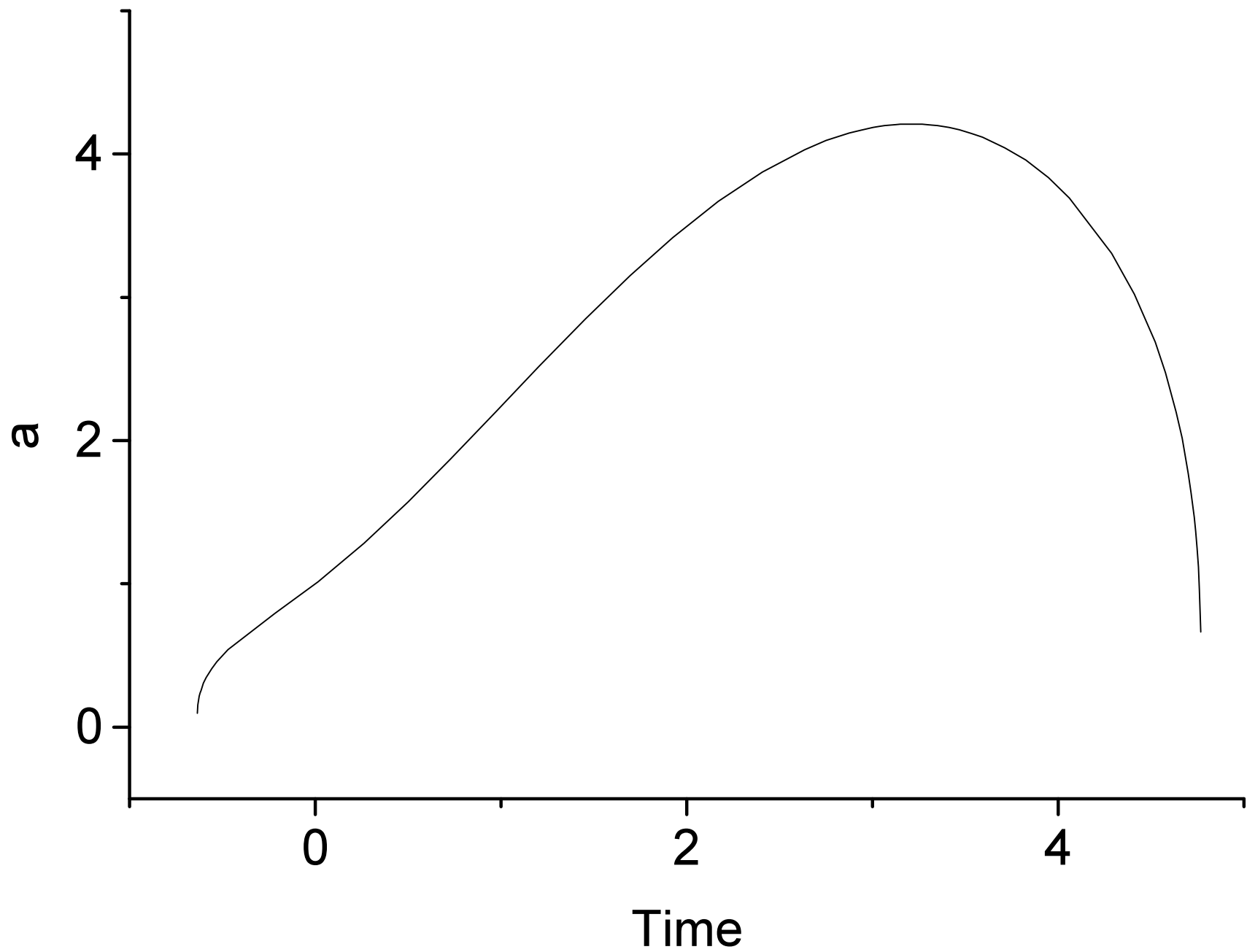
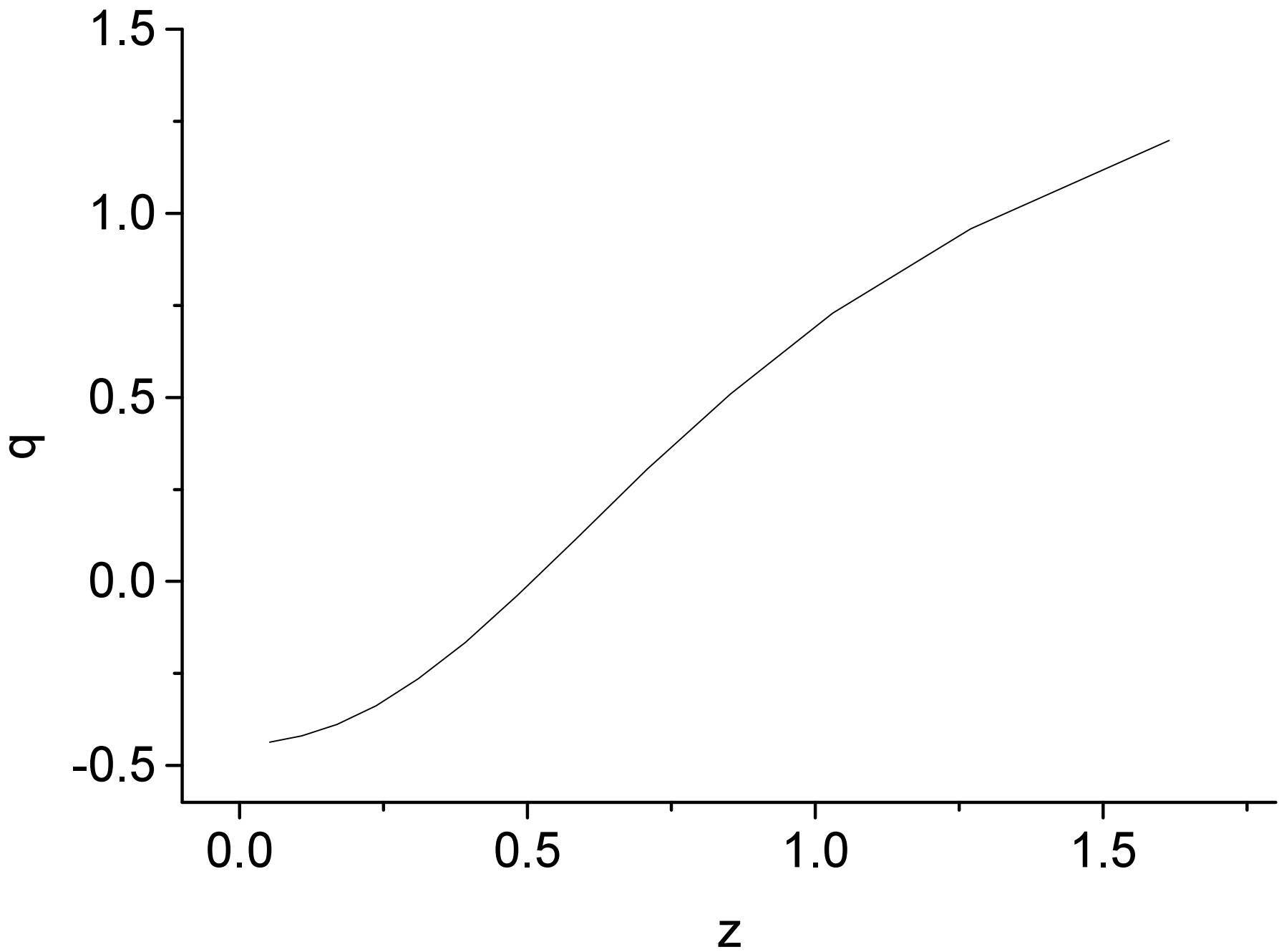
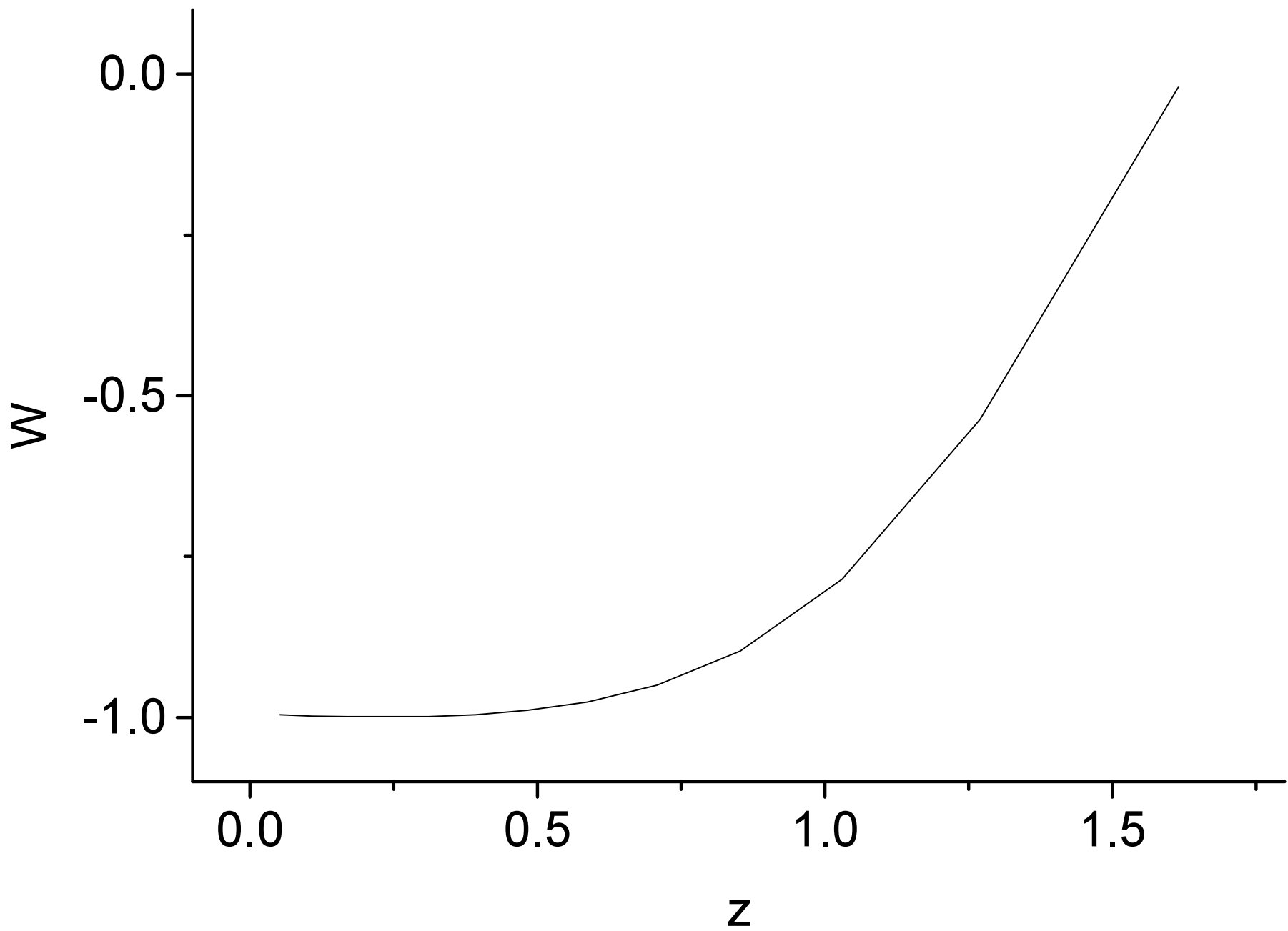
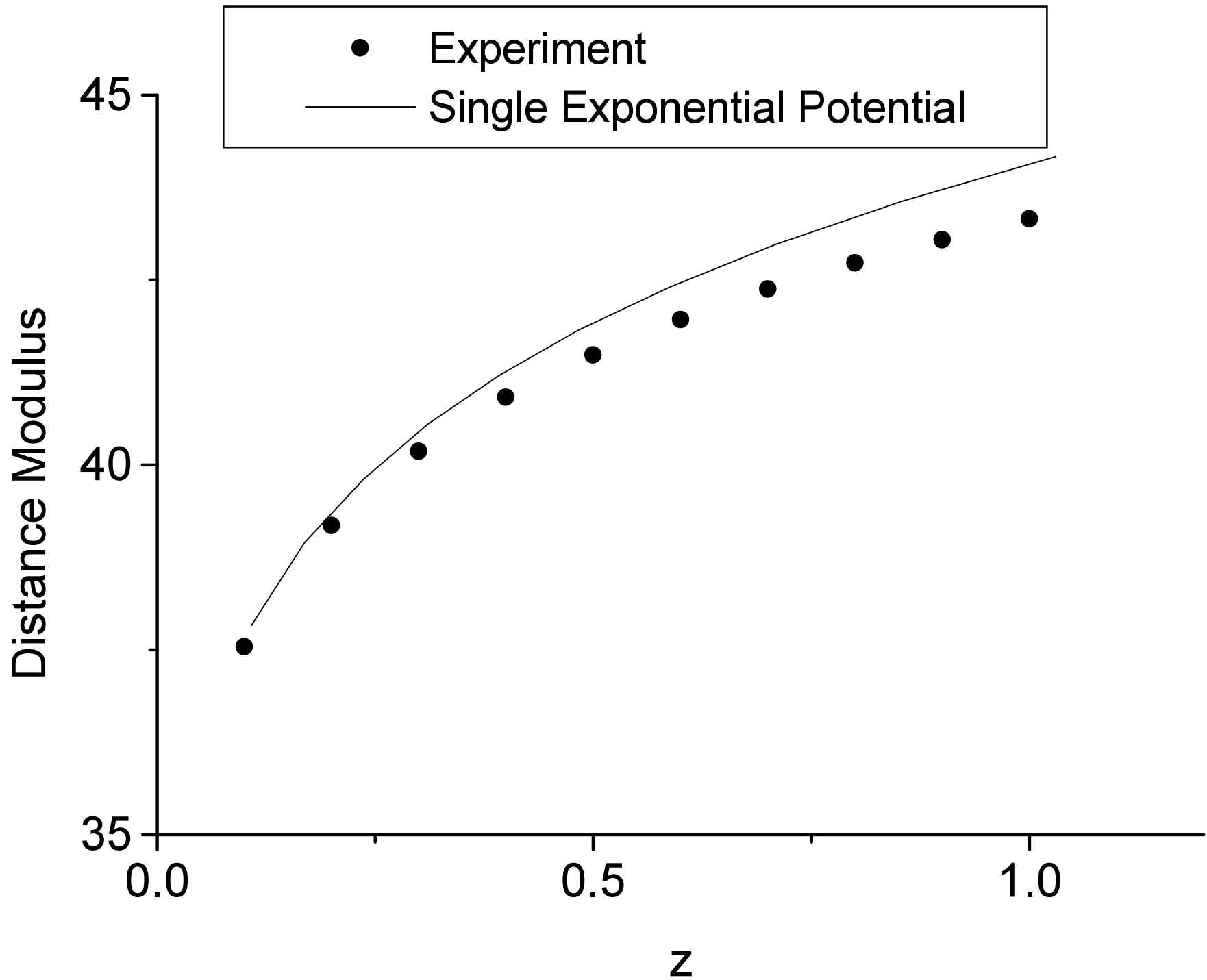
\end{document}